\documentclass{amsproc}
\usepackage[utf8]{inputenc}

\usepackage{graphicx}
\usepackage{amsmath}
\usepackage{amssymb}
\usepackage{amsthm}
\usepackage{bbm}
\usepackage{tikz}
\usepackage{tikz-cd}

\usepackage[maxbibnames=20,sorting=none, backend=biber]{biblatex}
\usepackage{caption}
\usepackage{subcaption}

\title{Semidirect Product Key Exchange: the State of Play }
\author{Christopher Battarbee, Delaram Kahrobaei, Siamak~F.~Shahandashti}
\dedicatory{Dedicated to the achievements of Joachim Rosenthal on the occasion of his 60th birthday.}

\begin{document}

\maketitle

\begin{abstract}
    Of the many families of cryptographic schemes proposed to be post-quantum, a relatively unexplored set of examples comes from \textit{group-based} cryptography. One of the more central schemes from this area is the so-called Semidirect Product Key Exchange (SDPKE), a generalisation of Diffie-Hellman Key Exchange that is plausibly post-quantum. In this report we survey the state of the literature relating to SDPKE, providing a high-level discussion of security, as well as a comprehensive overview of the proposed platforms and the main cryptanalytic ideas relevant to each.
\end{abstract}

\section{Introduction}
\subsection{Motivation}
Few fields possess a text as foundational as \textit{New Directions in Cryptography} \cite{diffie1976new}, which presents a key agreement mechanism today known as the Diffie-Hellman Key Exchange (DHKE). The protocol remains relevant in modern cryptographic applications and works as follows:
\begin{enumerate}
    \item Suppose Alice and Bob wish to establish a shared secret key $K$. They agree on a public, finite group $G$ and a generator $g\in G$.
    \item Alice picks a random integer $a$, which she keeps secret, and calculates $g^a$ (here the exponentiation refers to repeated application of the group operation). She sends this latter value to Bob.
    \item Bob similarly calculates group element $g^b$ from his secret random integer $b$ and sends this to Alice.
    \item Upon receipt of $g^b$ Alice uses her private exponent to calculate $K_A:=(g^b)^a$; similarly, Bob calculates $K_B:=(g^a)^b$. Since $g^{ab}=g^{ba}$ we have key agreement.
\end{enumerate}

Since all communication is assumed to be on an insecure channel, security relies on the hardness of the so-called \textit{Diffie-Hellman problem}: the recovery of $g^{ab}$ from the data $g, g^a, g^b$. The security is also related to the \textit{discrete logarithm problem} (DLP), recovery of $a$ from the data $g, g^a$, but the relationship is not fully understood. For discussion of the relationship between these problems and other problems in the class of Diffie-Hellman hardness assumptions, see, for example, \cite{katz2020introduction}.

Today, the security of DHKE is threatened by Shor's algorithm \cite{shor1994algorithms}, which is able to efficiently solve the hidden subgroup problem in at least finite abelian groups; i.e., the platform originally proposed for use with DHKE. Since there is a reduction of DLP to a hidden subgroup problem, we consider DHKE to be extremely vulnerable to quantum attack\footnote{The difficulty of DLP and the hidden subgroup problem in various groups is beyond the scope of this report: for a survey of the state of the hidden subgroup problem in various platform groups, see \cite{horan2018hidden}.}. To this end, the National Security Agency (NSA) announced plans in 2015 to upgrade security standards to so-called `post-quantum' protocols \cite{NISTPQ}. Indeed, after three rounds of selection, in July $2022$ this standardisation process bore its first fruit with the announcement of the third-round finalists, largely coming from the lattice-based and hash-based families of post-quantum cryptography \cite{dssannouncement}. Nevertheless, the standardisation process is ongoing into a fourth round, including a call for a more diverse roster of Digital Signature Schemes \cite{dssannouncement}. Moreover, two third-round candidates have in 2022 been the subject of high-profile attacks: the multivariate-based Digital Signature Scheme, Rainbow \cite{ding2005rainbow}, was shown to admit secret key recovery with the recommended parameter sets in \cite{beullens2022breaking}; and the isogeny-based Key Encapsulation Mechanism, SIKE \cite{jao2011towards}, was shown to admit secret key recovery in \cite{castryck2022efficient, maino2022attack}. These cases highlight the need to continue the cryptanalysis effort, and to increase the diversity of prospective post-quantum schemes. Indeed, the relatively unexplored field of \textit{group-based} cryptography is a prospective source of such schemes: the following is a survey of the proposed instances and cryptanalysis of one of the more well-known examples. 

\subsection{Beyond Diffie-Hellman: A New Key Exchange}
The current proposals for quantum-safe cryptosystems can be broadly grouped into six categories, one of which is known as \textit{group-based}. The proposal of our interest belongs to this category. In some sense, this category is the most natural extension of DHKE; indeed, many examples have DHKE as a special case. In our case, we appeal to a similar syntax to DHKE utilising a more complex group structure.

\subsubsection{The Semidirect Product}
Let $G,H$ groups and $\rho:H\to \mathrm{Aut}(G)$ a homomorphism into the automorphism group of $G$. The set $G\times H$ is a group, written $G\rtimes_{\rho}H$, when equipped with group operation defined\footnote{Other texts may define the semidirect product differently; in fact, the various notions are equivalent up to isomorphism. Indeed, by itself this is rather a shallow definition of the semidirect product; we are actually only using it to get a a more complex but efficiently calculable notion of exponentiation.} by 
\[(g,h)(g',h')=(g^{\rho(h')}g', hh').\]
If $H$ embeds in the automorphism group of $G$ then the homomorphism $\phi$ is just the identity, and the product simplifies to
\[(g,\phi)(h,\psi)=(\psi(g)h,\psi\phi),\]
where $\psi\phi:=\phi\circ\psi$ represents the function obtained by applying first $\phi$, then $\psi$. In this case we have an object known as the holomorph, written $G\ltimes \mathrm{Aut}(G)$: one verifies by induction that exponentiation takes the form
\[(g,\phi)^n=\left(\prod_{i=0}^{n-1}\phi^{n-(i+1)}(g),\phi^n\right),\]
where $\phi^n$ means the automorphism $\phi$ composed with itself $n$ times.

\subsubsection{Semidirect Product Key Exchange}
Armed with this machinery we define a key exchange mechanism known as semidirect product key exchange (SDPKE). The proposal in its full generality first appears in \cite{habeeb2013public}, although a revised version suggesting a new platform was later published \cite{kahrobaei2016using}, and works as follows. Suppose Alice and Bob agree on a public group $G$, as well as a group element $g$ and automorphism of $G$, say $\phi$, then they can arrive at the same $G$-element:
\begin{enumerate}
    \item Alice picks a random secret integer $x$, and calculates the holomorph exponent $(g,\phi)^x=(A,\phi^x)$. She sends \textbf{only} $A$ to Bob.
    \item Bob similarly calculates $(B,\phi^y)$ corresponding to random, private integer $y$, and sends only $B$ to Alice.
    \item With her private automorphism $\phi^x$ Alice can now calculate her key as the group element $K_A=\phi^x(B)A$; Bob similarly calculates his key $K_B=\phi^y(A)B$.
\end{enumerate}

Since 
\begin{align*}
    \phi^x(B)A &= \prod_{i=0}^{y-1}\phi^{x+y-(i+1)}(g)\prod_{i=0}^{x-1}\phi^{x-(i+1)}(g) \\
    &= \prod_{i=0}^{x-1}\phi^{x+y-(i+1)}(g)\prod_{i=0}^{y-1}\phi^{y-(i+1)}(g) \\
    &= \phi^y(A)B
\end{align*}

we have $K_A=K_B$. Note that the syntax is similar to that of classical DHKE; indeed, if our automorphism $\phi$ is the identity we have DHKE as a special case.

\subsubsection{Application to Encryption Schemes}
In \cite{elgamal1985public} a public-key encryption system based on the machinery of DHKE is proposed, now known as the ElGamal Cryptosystem. Analogously\footnote{Similarly to \cite{elgamal1985public}, a digital signature scheme is also proposed.} in \cite{moldenhauer2015group} a public-key encryption scheme is proposed, called the MR cryptosystem, which is detailed below. Note that exponentiation here refers to the semidirect product notion of exponentiation discussed above.
\begin{enumerate}
    \item Alice and Bob agree on a public group $G$, as well as a fixed element $g\in G$ and automorphism $\phi\in \mathrm{Aut}(G)$.
    \item Alice chooses a random secret integer $n$ and calculates public key $a$ via $(a,\phi^n)=(g,\phi)^n$.
    \item To send a message $m\in G$ to Alice using her public key, Bob computes $(c_1, \phi^r)=(g, \phi)^r$, where $r$ is a random ephemeral key, then computes $\phi^r(a)c_1$. The ciphertext will be the pair $(c_1,c_2)=(c_1,\phi^r(a)c_1m)$.
    \item Upon receipt of Bob's ciphertext Alice computes $K=\phi^n(c_1)a$ and decrypts by calculating $K^{-1}c_2$.
\end{enumerate}

We have correctness since $\phi^n(c_1)a=\phi^r(a)c_1$ (we show this in exactly the same way we demonstrate key agreement in the SDPKE protocol), so $c_2=Km$. Again, if the automorphism is the identity we have the classical syntax of ElGamal as a special case.

\subsection{Security of SDPKE}
For our purposes, we will consider an instance of SDPKE `secure' if one cannot efficiently recover the shared key $K$ efficiently provided access to public information and the values $A$ and $B$ exchanged in the clear. 

Consider the following computational problem: given $g,\phi$ and a product of the form $\prod_{i=0}^{x-1}\phi^{x-(i+1)}(g)$ for some $x\in\mathbb{N}$, recover the integer $x$. We call this task the Semidirect Discrete Logarithm Problem (SDLP). In the same way that the ability to solve the classic Discrete Logarithm Problem allows one to recover the shared key in the standard Diffie-Hellman key exchange, the ability to solve SDLP allows one to recover the shared key with respect to SDPKE.

The cryptanalysis reviewed in this survey does not achieve key recovery by solving SDLP. Instead, some underlying linearity of the platform is exploited to show that the shared key is leaked from public information. Nevertheless, recent work \cite{battarbee2021cryptanalysis} provides analysis of the computational problem SDLP, so before detailing the larger body of cryptanalysis we will briefly review the contents of \cite{battarbee2022subexponential}, and their implications for SDPKE.

The main result is as follows: fix a pair $(g,\phi)$. There is a value $r$ dependent on $(g,\phi)$ such that a quantum algorithm solves SDLP with respect to $(g,\phi)$ in worst-case time $2^{\mathcal{O}(\sqrt{\log r})}$. In particular, we have an upper bound on the quantum time complexity of solving SDLP, and therefore of achieving shared key recovery in SDPKE. The construction of this algorithm is established by drawing a surprising connection to group actions.

Recall that a group action consists of a tuple $(G,X,\star)$, where $G$ is a group, $X$ is a set, and $\star:G\times X\to X$ is a function. By convention, we write $\star(g,x)$ as $g\star x$. We require that $1\star x=x$; and that $(g\cdot h)\star x=g\star(h\star x)$ for all $g,h\in G$. One can use this machinery to come up with a generalisation of Diffie-Hellman Key Exchange as follows: suppose $G$ is finite and abelian, and that $X$ is finite. Indeed, suppose Alice and Bob both agree on a public group action $(G,X,\star)$ and some $x\in X$. They can arrive at the same $X$-element as follows:
\begin{enumerate}
    \item Alice picks $g\in G$ uniformly at random and calculates $X=g\star x$ and sends it to Bob.
    \item Bob picks $h\in G$ uniformly at random and calculates $Y=g\star x$, and sends this to Bob.
    \item Upon receipt of $Y$ (resp. $X$) Alice (resp. Bob) calculates $K_A=g\star Y$ (resp. $K_B=h\star X$).
\end{enumerate}

Correctness, i.e. the fact that $K_A=K_B$, follows from the definition of a group action and commutativity of the group. This framework originally appears in \cite{couveignes2006hard}\footnote{As a historical aside, note that this paper was written in 1997 but did not appear until 2006, following a renewed interest.}; a \textit{Hard Homogeneous Space} is a group action such that various useful operations are efficient, but the following problem is not: given $x,y$ sampled uniformly from $X$, find\footnote{Technically speaking, in order to guarantee the existence of such a $g$ it suffices to require that the action is both \textit{free} and \textit{transitive}. We will not delve into the definitions of these terms here, but we are assuming henceforth that all group actions we deal with are free and transitive.} $g\in G$ such that $g\star x=y$. Couveignes dubs this task the \textit{vectorisation problem}, but we will will refer to it as the Group Action Discrete Logarithm Problem (GADLP). Clearly, with the ability to solve GADLP one can break the group action-based key exchange defined above; so, if we know how to solve GADLP, and we can show that SDLP reduces to GADLP, we have an algorithm for solving SDLP.

In fact, we do know how to solve GADLP, at least with access to a quantum computer. It is reasonably well-known - see for example \cite{childs2014constructing} - that GADLP reduces to the Abelian Hidden Shift Problem, to which the important quantum algorithm of Kuperberg for the Dihedral Hidden Subgroup Problem  \cite{kuperberg2005subexponential} can be applied. It remains, therefore, to complete the group action reduction.

In order to accomplish this we show that an abelian group acts on a subset of the set of all products of the form $\prod_{i=0}^{x-1}\phi^{x-(i+1)}(g)$. It will be immediately convenient to write products of this form as $s(g,\phi,x)$ - indeed, when $(g,\phi)$ is a fixed pair we can think of the function $s$ as taking only integer arguments, analogously to the standard notion of group exponentiation. It turns out that the set $\{s(g,\phi,x):i\in\mathbb{N}\}$ - which must be finite, since it is a subset of a finite semigroup - is of the form 
\[\{g,...,s(g,\phi,n-1),s(g,\phi,n),...,s(g,\phi,n+r-1)\},\]
where $n,r$ are parameters determined as a function of $(g,\phi)$. The values $n,r$ are called the \textit{index} and \textit{period} of $(g,\phi)$ respectively; each of them define the \textit{tail} $\{g,...,s(g,\phi,n-1)\}$ and \textit{cycle} $\{s(g,\phi,n),...,s(g,\phi,n+r-1)\}$. Indeed, elements of the cycle are periodic in $r$ (hence the name), which allows one to deduce that a cyclic group $\mathbb{N}_r$ acts on the cycle. Now, if $n=1$, the tail is empty and we have that SDLP is exactly GADLP. This is not, however, generally the case, and to complete the reduction it will be necessary to extract the values $n,r$ from the pair $(g,\phi)$. This is done via canonical quantum period-finding methods adapted from \cite{childs2014quantum}, and so we are done.

% Finally we note that the notion of security defined at the outset of this section is effectively analogous to the Computational Diffie Hellman problem (CDH). We call our version the Semidirect Computational Diffie Hellman problem (SCDH). There is also a group action version of CDH - in recent landmark work \cite{montgomery2022full} it is shown that this problem is quantum equivalent to GADLP. We are able to extend this result to our case: in particular, one concludes that unrecoverability of the shared key in SDPKE is quantum equivalent to the hardness of SDLP, in contrast to the classical case. Unfortunately the methods of \cite{montgomery2022full} do not seem to be able to be extended to the relevant decisional problem, and so for the time being it remains out of reach to verify the quantum equivalence of SDLP and this stronger notion of security.

We conclude this section by noting two wider implications of the work described above. First, due to a recent landmark result of Montgomery and Zhandry \cite{montgomery2022full}, the reduction above allows one to conclude that SDLP and the security notion defined at the outset of this section are, in fact, quantum equivalent. We therefore have increased confidence in the quantum security of SDPKE, since unlike in the classical case we do not have to assume the difficulty of a related but distinct problem. Instead, security in this sense is quantum equivalent to a problem for which the best-known quantum algorithms are quantum sub-exponential - and attacks of sub-exponential complexity have been treated as tolerable in classical contexts.

We also point out that, as far as we know, this reduction implies that the semidirect product context is the second non-trivial example of a Hard Homogeneous Space as originally proposed by Couviegnes - the other major example comes from isogenies between elliptic curves. We therefore, in line with the naming conventions of this latter field, suggest a renaming of SDPKE to SPDH. SPDH stands for \textbf{S}emidirect \textbf{P}roduct \textbf{D}iffie-\textbf{H}ellman, and should be pronounced \textit{spud}.

\section{Proposed Platforms}
We here point out that if, in the construction of the holomorph discussed above, we allow the group $G$ to be a semigroup, then the construction $G\ltimes \mathrm{Aut}(G)$ is also a semigroup. Many of the proposed platforms are in fact semigroups. Moreover, if $H$ is a semigroup and $\rho$ is a homomorphism into the endomorphism group of the group $G$, the construction $G\ltimes_{\rho}H$ is also a semigroup.

\subsection{Matrices over Group Rings}
The authors of \cite{habeeb2013public} propose the platform semigroup $M_3(\mathbb{Z}_7[A_5])$ with automorphism defined as conjugation by some invertible matrix in the semigroup. Here, $\mathbb{Z}_7[A_5]$ denotes the group ring consisting of formal sums of the form 
\[\sum_{g\in A_5}a_g.g\quad a_g\in\mathbb{Z}_7.\]
One can define a notion of addition and multiplication on this ring; equipped with these operations we get a ring that is at the same time an algebra over an $|A_5|$-dimensional vector space over $\mathbb{Z}_7$. In this case we get a closed form of the security assumption; if $M$ is the public semigroup element and $H$ is the matrix defining the conjugation, security is reduced to the problem of retrieving the quantity $H^{-(m+n)}(HM)^{m+n}$ from the data $H, M, H^{-m}(HM)^m, H^{-n}(HM)^n$: one must in this case be careful that the matrices $H$ and $M$ do not commute.

In \cite{moldenhauer2015group} we use a similar automorphism; however, since we need invertibility for decryption we use platform group $GL(r,K)$ for some field $K$. No specific parameters are suggested; we note that throughout the literature $r=3$ seems standard. The security assumptions made take the same form as those in \cite{habeeb2013public}.

\subsection{Free Nilpotent p-Groups}
In \cite{kahrobaei2016using} the authors, addressing security issues in \cite{habeeb2013public} to be addressed later on, propose nilpotent $p$-groups as a platform. Of course, all finite $p$-groups are nilpotent (see, for example, \cite[Theorem~5.33, p.115]{rotman2012introduction}); here we explicitly generate them via factor groups of free groups; for example, consider the free group on $r$ elements $F_r$. The normal subgroup $F_r^p$ is generated by all elements of the form $g^p$ for $g\in F_r$ and some prime $p$. With notation $[a,b]$ denoting the product $a^{-1}b^{-1}ab$ and $[[a_1,...,a_n],a_{n+1}]$ defined inductively, the normal subgroup generated by all elements of the form $[[a_1,...,a_{c-1}], a_c]$ is denoted $\gamma_{c+1}(F_r)$. Assembling these components the finite group $F_r/F_r^{p^2}.\gamma_c(F_r)$ is proposed for the platform
This is done to enforce a low nilpotency class for efficient calculation, and to ensure the existence of an element of order $p^2$, which as we will see later has useful security properties. The authors note that efficiency seems to depend on the values of $c$ and $r$, which can safely be kept low if we use a very large prime $p$.

% In fact some of these groups occur naturally as semidirect products; for example, non-abelian groups of order $p^3$ for an odd prime $p$ - see \cite{conrad} for useful exposition on the classification of these groups.

\subsection{Tropical Algebras}
A tropical semiring is a subset of $\mathbb{R}$, containing $0$, and equipped with a notion of addition (written $\oplus$) and multiplication (written $\otimes$) defined by 
\[x\oplus y = min(x,y) \quad x\otimes y = x+y.\]
Consider matrices with entries in this semiring with the obvious component-wise definitions of addition and scalar multiplication. We can also multiply two matrices by replacing the operations in the usual definition of matrix multiplication with the semiring operations; it turns out this gives us a module over a semiring equipped with a bilinear product, which we will call the \textit{tropical algebra}.

In \cite{grigoriev2014tropical} a key exchange over this algebra was proposed; it was broken in \cite{kotov2018analysis}. The authors therefore update their protocol to include the SDPKE syntax in \cite{grigoriev2019tropical}, using an operation called \textit{adjoint multiplication}, defined by $a\star b = a\oplus b\oplus (a\otimes b)$. Unlike its analogue in the usual arithmetic this operation is distributive, so the action of the algebra on the semigroup formed by the algebra considered under addition by $H\cdot G=G\star H$ is a semigroup action, and we can use it to define a semidirect product structure. The aim is to present a key exchange that is extremely efficient, since no multiplication is involved, and not vulnerable to the attacks against the above two schemes. A further key exchange corresponding to a public endomorphism induced by a similar action is also proposed; the two schemes are closely linked.

\subsection{Matrix Action Key Exchange (MAKE)}
In \cite{rahman2022make} the authors consider the set of $n\times n$ matrices over a finite field $\mathbb{Z}_p$. This object is a monoid under the standard notion of matrix multiplication and a group under matrix addition. Call this group $G$; in fact, restricting to the semigroup $S=\{(H_1^i,H_2^i):i\in\mathbb{N}\}$ where $H_1,H_2\in\mathbb{Z}_p$ are non-invertible, the action of $S$ on $G$ defined by $(H_1^i,H_2^i)\cdot M=H_1^iMH_2^i$ induces a homomorphism into the endomorphism group of $G$, so we get a semidirect product of $G$ by $S$ with exponentiation
\[(M,(H_1,H_2))^n=\left(\sum_{i=0}^{n-1}H_1^iMH_2^i, (H_1^n,H_2^n)\right).\]

The authors are able to show that recovery of the private exponent in a transmitted value is as least as hard as private exponent recovery in classical DHKE, and posit that the improved mixing from the combination of operations is beneficial for security. Moreover, attacks that threatened other instances of the scheme do not seem to directly apply.

\subsection{Matrices Over Bit Strings (MOBS)}
The authors of \cite{rahmanmobs} propose a platform of the holomorph of matrices over a semiring, serving two purposes: following \cite{grigoriev2014tropical} we use a semiring to mitigate some of the damage done by powerful representation-type attacks, and we address a weakness of \cite{rahman2022make}. The semiring is bitstrings where addition is given by bitwise OR and multiplication by bitwise AND; the automorphism is constructed just by permuting each bitstring in a matrix, where we use prime-order cycles to derive a high-order permutation.

\section{Main Attacks}

In general attacks against SDPKE fall into two broad categories, which we will detail below; however, we first turn our attention to the attacks on the tropical cryptography, which do not fit into either.

\subsection{Cryptanalysis of the Tropical Cryptography}
There are two attacks against the tropical cryptography, both of which achieve recovery of the private exponent by analysing the sequence of possible exchange values, which we will write $\{M_n\}$. In \cite{rudy2021remarks} the authors prove that, relative to a partial order defined on the matrix algebra, the sequence $\{M_n\}$ is monotone decreasing. Leaving aside some minor subtlety this effectively allows recovery of the private exponent via a simple binary search. The idea of \cite{isaac2021closer} is similar: the authors use the fact that the above sequence of matrices has a property known as \textit{almost linear periodicity} to extract the private exponent. In this case there is a small chance of algorithm failure, but the authors provide experimental evidence the the attack method is on average faster. In both cases, once the private exponent, say $x$, corresponding to an exchange value $A$, the key can be calculated from the other exchange value $B$ via $K=\phi^x(B)A$.

\subsection{A Word on Representations}
The strategy of many attacks against SDPKE schemes will be to construct a representation from the proposed platform group, which is by design obfuscating and not often well-understood, into a more familiar space. For our purposes, a faithful representation of a group is an injective homomorphism into $GL_n(K)$ for some field $K$. Similarly, a faithful representation of a semigroup is an injective homomorphism into $M_n(K)$ for some field $K$. These quantities will be presented as matrices depending on a particular (and usually arbitrary) choice of basis of the $n$-dimensional vector space over $K$; this value $n$ is known as the dimension of the representation. As a consequence of Cayley's theorem (and the analogue available for semigroups) all the objects we consider will admit such a representation, but such a representation may only be available for certain choices of $n$; we can therefore talk about the efficiency of a representation depending on the minimum size of $n$ for which we can construct a faithful representation.

\subsection{The Dimension Attack}

In \cite{roman2015linear}, following work in \cite{myasnikov2015linear}, the authors demonstrate that for SDPKE over a platform $G\ltimes \mathrm{Aut}(G)$ with public parameters $g, \phi$, one can compute the shared secret key directly using only public information if $G$ is a multiplicative subgroup of a finite dimensional algebra over a field $K$, and the endomorphism $\phi$ is extended to the underlying vector space $V$ of the algebra. By conditions in \cite{myasnikov2015linear}, provided operations in the underlying field $K$ are efficient, we can efficiently find a finite maximal linearly independent subset of the set $\{a_0,a_1,...,a_k,...\}$, where $a_i=\prod_{i=0}^{n-1}\phi^{n-(i+1)}(g)$ are the exchange values corresponding to all possible private exponents. Note that in this notation the exchange values $A,B$ are such that $A=a_x, B=a_y$, and the key $K$ is such that $K=a_{x+y}$. Suppose $\{a_0,...,a_k\}$ is this maximal linearly independent subset. It is shown that one can find coefficients $\eta_i$ such that
\[a_x=\sum_{i=0}^k\eta_i a_i,\]
so that, since $\phi$ is extended to an endomorphism of the algebra, we have
\begin{align*}
    K=\phi^y(a_x)a_y &= \phi^y\left(\sum_{i=0}^k\eta_i a_i\right)a_y \\
    &= \sum_{i=0}\eta_i\phi^y(a_i)a_y = \sum_{i=0}\eta_i\phi^i(a_y)a_i
\end{align*}
Since all the data in this latter sum is known to the attacker we have achieved key recovery. We will refer to this strategy as the \textit{dimension attack}.

For our purposes the algebras in question will almost always be a finite matrix algebra; that is, matrices which we can add, scale and multiply. In this case finding the above coefficients amounts to solving $n^2$ linear equations where $n$ is the size of the matrix, which will dominate the complexity of the algorithm.

It is for this reason we are interested in faithful representations of the various platforms - it allows us to embed the platform as a multiplicative subgroup of a matrix algebra and thereby carry out the dimension attack. In the case of \cite{habeeb2013public}, the candidate platform $M_3(\mathbb{Z}_7[A_5])$ is already a matrix algebra, so in a sense we consider a trivial, identity representation. We conclude that $M_3(\mathbb{Z}_7[A_5])$ is not a good choice of platform for the scheme, since it is a low-dimension matrix algebra and we expect the dimension attack to perform efficiently.

The situation is more complicated for the platform suggested to address these issues in \cite{kahrobaei2016using}. This is because the platform group is no longer a matrix algebra by default so we must consider a non-trivial representation into a space where the dimension attack applies. In \cite{janusz1971faithful} it is shown that $p$-groups with at least one element of order $p^n$ for some $n\in\mathbb{N}$ have lower bound $1+p^{n-1}$ on the dimension of admissible faithful representation in the requisite matrix algebra. 

Recall that in \cite{kahrobaei2016using} a group with an element of order $p^2$ is proposed. The group is defined on an alphabet $\{x_1,...,x_r\}$ for each $r\in\mathbb{N}$, and is moreover defined such that one can artificially control its nilpotecy class $c$; fixing $c=2$, each element of the group has the form 
\[x_1^{\alpha_1}...x_r^{\alpha_r}[x_1,x_2]^{\beta_{1,2}}...[x_i,x_j]^{\beta_{i,j}}...[x_{r-1},x_r]^{\beta_{r-1,r}}\]
where $[x_i,x_j]$ is the commutator of $x_i$ and $x_j$, and $\alpha_i$ and $\beta_{i,j}$ each range between $0$ and $p^2$. It follows that the group has order $(p^2)^r(p^2)^{r(r-1)/2}=p^{r^2+r}$, and can thus be represented by bitstrings of length $(r^2+r)\log p$. Our security parameter - that is, the length of the public keys - is therefore a function of $r$ and $p$.

Meanwhile, the dominating factor in terms of complexity in \cite{roman2015linear} is Gaussian elimination. We use the standard estimate that Gaussian elimination in an algebra of dimension $n$ over a field has complexity $\mathcal{O}(n^3)$, and therefore estimate that the dimension attack against the group discussed above has complexity about $\mathcal{O}((p+1)^3)=\mathcal{O}(p^3)$. Writing $k=(r^2+r)\log p$, in order to estimate the complexity of the attack in terms of a security parameter it remains to find a function $f$ such that $f(k)=\mathcal{O}(p+1)^3$. With the lowest possible choice of $r=2$ one has $2^k=2^(6\log p)=p^6$ - in other words, with $k$ the security parameter, the dimension attack runs in time $\mathcal{O}(\sqrt{2^k})=\mathcal{O}(2^{k/2})$. More generally, the attack runs in time $2^{\mathcal{O}(k)}$, where the constants get worse with increasing $r$.

This is the useful security property of the proposed free nilpotent $p$-groups alluded to earlier; the authors of \cite{myasnikov2015linear} point out that it remains an interesting question to characterise the efficiency of suitable representations of various linear groups to give a picture of suitable choices of platform group for SDPKE.

Finally, we make the rather obvious, though as far as we know original, observation that this attack will also break the MR cryptosystem proposed in \cite{moldenhauer2015group}. This is because one has $c_2=Km$, so one only needs to recover the blinding `key' factor $K$ to allow for complete message recovery. Since $K=\phi^n(c_1)a$ and $a,c_1$ are transmitted in the clear, we can think of these values as the exchange values in the SDPKE protocol and apply the same attack.

\subsubsection{The Non-linear Decomposition Attack} The authors of the dimension attack have also referred to their method as a `linear decomposition attack', in reference to the linearity exploited when embedding a multiplicative group into an algebra. In response to the results that this cannot always be done efficiently, a \textit{non}-linear decomposition attack is introduced, which does not rely on such an embedding - instead, finitely many generators for the subgroup generated by all possible exchange values are recovered by means of the ability to solve the membership problem in the group $G$. Moreover, it is shown that for essentially the same reasons that key agreement holds, knowledge of such generators suffices to recover the shared key $K$.

It appears that the specific low-nilpotency parameters suggested in \cite{kahrobaei2016using} succumb to this approach, and indeed it is known that the membership problem is feasible in all nilpotent groups. Moreover, we have seen that $p$-groups are a useful tool to tackle the vulnerabilities exposed by the linear decomposition attack, and all finite $p$-groups are known to be nilpotent.

On the other hand, numerous questions remain. We would like to investigate what happens when the nilpotency class is varied, the size of the subgroups generated, and how the complexity grows as a function of the security parameter, and so on.

\subsection{The Telescoping Equality}
\subsubsection{Cryptanalysis of MAKE}
As well as the improved mixing posited by the authors of MAKE \cite{rahman2022make}, the platform group is an additive subgroup of a matrix algebra and so does not naturally embed as a multiplicative subgroup as per the requirements of the dimension attack. There has therefore been an effort to develop alternative methods of key recovery; it turns out this new strategy is similar to the dimension attack in that it bypasses the proposed hardness assumptions.

Remarking upon an earlier version of MAKE in which $H_1=H=H_2$ and $H$ is invertible, the authors of \cite{monico2020remark} are the first to point out the identity 
\[H^nAH^n+\sum_{i=0}^{n-1}H^iMH^i=HAH+M.\]
However, the method of key recovery proposed relies on the invertibility of the matrix $H$. Following this work and the update to the current version of MAKE \cite{rahman2022make}, the authors of \cite{brown2022cryptanalysis} achieve key recovery by noting that 
\[H_1AH_2+M=H_1^xMH_2^x+A.\]
All terms in this equation can be calculated by an eavesdropper except $H_1^xMH_2^x$, which in this case can be uniquely recovered since we have an additive inverse. Upon recovery of this quantity one uses a consequence of the Cayley-Hamilton theorem to prove the existence of a vector $s$ dependent only on $H, x$ that solves the equation
\[L(Y)s=vec(H_1^xMH_2^x)\]
for any $Y\in M_n(\mathbb{Z}_p)$, where $L$ is a function defined by the authors, and $vec$ stacks columns of a matrix to get a vector. Using known data we can solve the above system of equations to get a vector $t$, so defining $u=t-s$ we must have $L(M)u=0$. By a technical intermediary lemma and the fact that $L$ is an additive homomorphism we have that $L(B)u=0$, so $L(B)t=L(B)s=vec(H_1^xBH_2^x)$. Since $K=H_1^xBH_2^x+A$ and $A$ is known to an eavesdropper we conclude that, since the complexity of the attack is dominated by the polynomial time process of recovering the vector $t$, there is a polynomial time algorithm to achieve key recovery.

In fact, even in some situations where the Cayley-Hamilton theorem does not apply we are able to achieve key recovery, again by constructing a representation \cite{battarbee2021cryptanalysis}. A notable example of a platform for which the Cayley-Hamilton theorem does not immediately apply but key recovery is still possible are group rings of the type used in \cite{habeeb2013public}; such group rings, having a group as a basis, can be embedded in a matrix algebra over a commutative ring. It turns out replacing the group ring entries of each matrix with their matrix representations gives a ring homomorphism, so we can embed injectively into a matrix algebra over a commutative ring and carry out the attack using Cayley-Hamilton in this latter space. Note that in contrast with the dimension attack, in this case the goal of the representation is not to embed the platform as a multiplicative subgroup, but to embed the platform into a space where the Cayley-Hamilton theorem applies.

\subsubsection{The Telescoping Attack}
The authors of \cite{brown2022cryptanalysis} are basically exploiting a particular instance of the general fact relevant to all SDPKE schemes that
\[\phi(A)g=\phi^x(g)A.\]
We call this equation the \textit{telescoping equality}; it arises by splitting up a certain product in two different ways. Suppose some eavesdropping party has observed one round of the protocol; the data $g,\phi$ and $A$ are all available to this party, and therefore $\phi(A)$ can be calculated; the quantity $\phi^x(g)$ may be therefore be recovered by the telescoping equality. 

Intuitively, the quantity $\phi^x(g)$ appears to encode information about the exponent $x$. In fact, slightly more is true; note that certainly if we can recover the automorphism $\phi^x$ we can calculate $\phi^x(B)A=K$ and achieve key recovery. In fact, by arguments in \cite{monico2021remarks} we also have that any $q\in\mathbb{N}$ such that $\phi(A)g=\phi^q(g)A$ has $\phi^q(B)A=K$.

We will refer to this general strategy as the \textit{telescoping attack}. An attacker following this strategy is required to overcome two main problems: recovery of a value quantity $\phi^q(g)$ for which $\phi(A)g=\phi^q(g)A$, and use of this quantity to recover information about the secret key.

The telescoping attack succeeds against the MAKE scheme (which we are now writing as in the general case) partly because we do not have to worry about non-unique solutions to the telescoping equality. This is because the platform object is a full group, and the action of a group on itself by left multiplication is transitive; that is, if we know the quantity $\phi^x(g)$ satisfies $\phi(A)g=\phi^x(g)A$, then any $Y$ satisfying $\phi(A)g=YA$ must be such that $Y=\phi^x(g)$. Indeed, this line of argument applies to all groups; in a semigroup there are possibly several admissible values of $Y$ in the equation $\phi(A)M=YA$ - and in particular, values admissible in this equation that are not equal to $\phi^q(g)$ for some $q\in\mathbb{N}$.   

To address the second problem we note also that recovery of $\phi^x(g)$ in the MAKE scheme essentially allows recovery of $\phi^x(B)$ for free. This is not the case in general; the function $L$ defined in \cite{brown2022cryptanalysis} is not a multiplicative homomorphism, so the strategy will only work against additive schemes. Moreover, the existence of a constant vector solving an equation for all platform elements is crucial, and its existence is derived through the Cayley-Hamilton theorem which only applies to a few of the proposed platforms. For other schemes the best strategy seems to be to recover the endomorphism $\phi^x$ from $\phi^x(g), g$, which gives us a reduction to semigroup DLP in the endomorphism semigroup of $G$. It is known that the semigroup DLP is essentially no worse than the classical DLP \cite{banin2016reduction}, \cite{childs2014quantum}.

\subsubsection{On the cryptanalysis of MOBS}
Before the results in \cite{monico2021remarks} were made available it was thought that only recover of the value $h^x(M)$ sufficed for key recovery - that is, it was not known that there are integers $q\neq x$ such that one can achieve key recovery by the same method with access to $h^q(M)$. As a result it was thought that the telescoping attack would have to proceed according to the following strategy: recover $h^x(M)$ from the equation $h(A)M=YA$, and use $h^x(M)$ to recover the key $K$. Since a general method of recovering $h^x(M)$ was not known, in \cite{battarbee2021efficiency} investigation was made into the number of $Y$ satisfying the equation $h(A)M=YA$ in order to investigate the feasibility of a `naive' attack of recording all solutions and trying each of them. In the paper, computational evidence is presented that even for relatively modest parameters, there are far too many solutions for this to be viable.

On the other hand, in \cite{monico2021remarks} a successful attack against the MOBS cryptosystem is presented. The attack exploits the decomposition of the automorphism into low-order cycles and in particular does not depend on an artefact of MOBS itself; that is, there are feasibly other versions of MOBS with differently assembled automorphisms that resist this attack. For these distinct choices of automorphism, say $\psi$, it would be interesting to update the work of \cite{battarbee2021efficiency} and compute the number of admissible values of $Y$ in the equation $\psi(A)M=YA$ such that there exists $q\in\mathbb{N}$ with $Y=\psi^q(M)$.

\section{Conclusion}
In summary the SDPKE protocol is promising but still faces challenges. Perhaps the most plausible looking schemes are MOBS, should its public automorphism be tweaked, or certain classes of $p$-group admitting only inefficient representations. In the general case we have a quantum reduction of the underlying security problem to, more or less, the Dihedral Hidden Subgroup Problem, which is not currently considered to be quantum-vulnerable. For the time being, since no other reductions are known, we have confidence in the viability of SDPKE as a post-quantum alternative to DHKE. However, as we have seen, there are numerous platform-specific security deficiencies yet to be addressed; much further study in this area is therefore required. 

\textbf{Acknowledgement}
We wish to thank Vladimir Shpilrain for reading this manuscript and provided helpful comments. We would also like to thank an anonymous reviewer for their thoughtful and detailed comments, according to which we have made several improvements to this manuscript.

\printbibliography

\end{document}